\magnification=1200
\vsize=8.1truein
\hsize=5.9truein
\baselineskip=20truept
\parskip=4truept
\vskip 18pt
\def\today{\ifcase\month\or January\or February\or
March\or April\or May\or June\or July\or
August\or September\or October\or November\or
December\fi
\space\number\day, \number\year}
 
\centerline{\bf SIZE DEPENDENCE, STABILITY, and the TRANSITION to BUCKLING} 
\centerline{\bf in MODEL REVERSE BILAYERS. }
\medskip
\centerline{by }
\centerline{ J. Stecki }
\centerline{Department III, Institute of Physical Chemistry}
\centerline{ Polish Academy of Sciences }
\centerline{ ul. Kasprzaka 44/52, 01-224 Warszawa, Poland}
\bigskip
\centerline{\today} 
\vskip 60pt

\centerline{Abstract}
Molecular Dynamics simulations of a model bilayer made of surfactant 
dimers in a Lennard-Jones solvent are reported for three sizes of the 
systems up to an area of $100\sigma \times 100\sigma$ and for a large 
interval of the specific areas: 
from hole formation under tension to the floppy 
state of a buckling compressed bilayer. The transition  to the floppy state
appears quite abrupt and discontinuous;  in the floppy state the 
lateral tension is negative.\hfill\break
Lateral tension and the structure factor were determined for all 3 sizes and all areas;
 the apparent rigidity constant and apparent surface tension
are determined and correlated with the specific area and the finite size.
 The replacement of the $1/q^2$ capillary-wave divergence by a pole is 
accounted for and explained.

\vfill\eject

\noindent {\bf I. INTRODUCTION. }

The selfassembly of amphiphilic molecules dissolved in a solvent, leads
to formation of bilayers. The properties of these two-dimensional sheets 
are of great interest because of their role in living matter. But 
it is also of interest to investigate the conditions for their formation and 
equilibrium existence as well as the limits of their stability. In particular,
the  liquid bilayer need not be necessarily 
formed by long-chain molecules; bilayers have been obtained in simulations
 with amphiphilic chains 4 segments long[1,2] and even have been formed by 
dimers - the shortest chain possible
[3]. Also, trimers in vacuum have been shown to form a stable flat sheet of 
a bilayer[4]. In [5] we obtained bilayers formed by amphipilic 
dimers, of a new kind, namely "reverse" bilayers, so named by analogy 
with reverse micelles. Though it may appear that a bilayer can be very 
easily formed but that is not so: in each of these systems the bilayer
is formed only within  limits of temperature, density, and  for given
interparticle and intermolecular interactions. Not much about these limits
is known.
Always in the same system micelles
of various shapes can be formed, or the amphiphilic molecules may 
disperse as a solution in the liquid solvent, if the thermodynamic state
favors either. Thus the limits of existence of bilayers and their stability 
are of interest. Moreover, we   have briefly reported[6] for 
bilayers made of chain molecules unexpected discontinuties in the 
transitions between the extended and floppy states and it is of interest 
to check if such discontinuities also appear in the bilayers made of dimers.
Those findings[6] for length $l=8$ segments were confirmed since 
for $l=4$[7].
  
 In this paper our previous work on bilayers formed by 
 the shortest chain-molecule possible, the dimer[5], is extended  
to much bigger systems in order to  study the size dependence. 
 The size dependence study includes the structure factor $S(q)$, 
 describing the shape fluctuations of the bilayer.  The  
 presence of regions of the wave-vector $q$ for which $S(q)$ would have to
 be negative, and the related disappearance of the capillary wave
divergence, is explained and resolved.

 In Section II the rather extensive results on the lateral 
tension are reported and discussed. The transition to floppy buckling 
bilayer is also examined.  The example of the apparent discontinuity in the 
intermolecular energy is  given.
 In Section III - the results on the structure factor are given.
 Section IV is devoted to  discussion and summary.

 The details of the model and the parameters
 of the simulations are given in  Appendix A.
\bigskip
\medskip

\noindent {\bf II. THE LATERAL TENSION AND THE SIZE DEPENDENCE }.

The lateral tension $\gamma$ of a bilayer depends on its area $A$.
As the bilayer is contained in a square box of volume $V=L_zL_x^2$ with 
periodic bouudary conditions, the only area available for external control is 
$A \equiv L_xL_y=L_x^2$, the edge of the box. $A$ is often called the 
"projected area".
 The quantity  $\gamma$ is the free energy increase due to area increase at
constant volume $V$, $\delta F = \gamma \delta A$. It is  
computed by the same molecular 
formula as the surface (interfacial) tension between liquid and vapor 
(or any other two fluid phases) - {\it i.e.} by the Kirkwood-Buff 
formula[1-7,8,9] as rederived for deformations of a parallelepiped[9]. 
Constant temperature $T$ and particle numbers $N, N_d$ being understood,
$$ \gamma \equiv (\partial F/\partial A)_V~.   \eqno(2.1) $$ 
However, the properties of $\gamma$ are very much unlike 
those of the surface tension and, partly for that reason, it is 
called the lateral tension.
The specific area per surfactant head is defined as
$$ a \equiv  2A /N_d~,  \eqno(2.2) $$ 
where $N_d$ is the number of surfactants. 
Not only does $\gamma$ depend on $a$, but also   
it can take negative values or zero; the state (i.e. the particular area
$a_0 = 2 A_0/N_d$)
at which $\gamma=0$, is the "tensionless state".  
In view 
of the definition (2.1), at the tensionless state 
$$  \partial F/\partial A \vert_{A=A_0} =0. \eqno(2.3)  $$
Stability requires (at constant $T,V,N$ again) 
$$   d\gamma /da > 0 \eqno(2.4)$$
or
$$ \partial^2 F/\partial A^2~>~0. \eqno(2.5) $$ 
This derivative is related to bilayer lateral (inverse) 
compressibility[1], $K_A$, formally defined as
$$ K_A=a(\partial \gamma/\partial a)  > 0~~~.\eqno(2.6) $$
It must be positive.
But $\gamma$ itself can be positive, null, or negative; at
positive $\gamma$ the slope of $F$ favors smaller areas (the familiar 
situation for  interfaces with surface tension) whereas for negative 
$\gamma$ the bilayer prefers to expand towards the tensionless state, (which
is the state of lowest free energy, at least locally).

In our previous simulation work[5] we have produced 
the values of $\gamma$ and thus the function $\gamma (a)$
in series of simulations with different areas, at constant volume and at the
same thermodynamic state.
 Unlike in earlier investigations[1,2,3,10], we explored[5] negative lateral 
tensions along with the usual zero and positive tensions in a  
range of areas as large as possible,
 i.e. often within the entire range of stability of the 
bilayer.  The results produced[5] curious and unusual shapes of 
the curves $\gamma (a)$. Also, in the transition range (near the tensionless
area $a_0$) there were hints of inflexions in the $F(a)$ curves obtained by 
numerical integration. It was therefore imperative to find which of these
new features would be present in larger systems, and to study 
quantitatively the size dependence 
by extending our earlier work to much larger systems.  
                                                             
 Fig. 1 shows the three plots of $\gamma(a)$ for three systems of 
nominal sizes $L_x\sim 36$, $L_x\sim 55$, and $\L_x\sim 100$ (the unit 
of length being $\sigma$, the collision diameter). 
 Hence all areas are in units of $\sigma ^2$.
The units used and the parameters are given in all detail in Appendix A. 

 Remarkably,  positive $\gamma$'s fall on a common curve; there
is no visible effect of the size. 
For small areas another region is obtained with a very small,
almost negligible, positive slope and a negative $\gamma$.
 In the latter region a 
clear size dependence can be seen: smaller $\vert \gamma \vert$ for 
larger system. The density profiles, i.e. the $z-$dependence of concentrations,
and the direct imaging of the system spatial configuration, show 
 a kind of a rough, floppy, fuzzy, 
crumpled, foamy, or buckled, state of the bilayer in this region.
Often the bilayer is full of solvent particles which suddenly 
find it possible to penetrate into the floppy bilayer.

The conclusion is that there are two regions, (E) and (C).
In (E), the region of large specific areas $a$, the bilayer is 
extended and gently undulating, the system is stable, $\gamma>0$,
the slope $d\gamma/da >0$ is positive, large, and common to all sizes.
 No visible size dependence in 
$\gamma(a)$ is seen as the data points fall on a common curve. 
In (C), the region of small specific areas $a$, the bilayer is 
compressed and floppy, the lateral tension is negative
$\gamma<0$. The derivative $d\gamma/da >0$ 
throughout the entire range. This implies stability. 
 In  few instances, at the very lowest investigated values of $a$, 
 the sign of the derivative was uncertain in
the largest system, still practically vanishing within the 
scatter of the data points.
 
 At each size there is a smooth transition between  (E) and  (C), 
 as can be seen in Fig.1. This transitional region is rather narrow.

The scaling with size of negative  tension data  
in region (C), was found surprisingly unambiguous.

The area dependence in (C) is very weak;
a  plateau value ${\gamma}_p$ of $\gamma$ can be assigned to each size. 
When plotted  against the system size, a very satisfactory 
scaling is obtained in Fig.2. For the measure of the size one can take 
the edge $L_x=L_y$, or its square, the area $A=L_xL_y$, or 
$N_d=2A/a$.  
Clearly  ${\gamma}_p$ tends to zero. Moreover it does so along a
straight line ${\gamma}_p = -const./A$.
 
   There is one calculation[11] which explicitely predicts a negative 
lateral tension 
$\gamma$ and its saturation to the lower bound given by 
$$ \kappa q_{min}^2 +\gamma = 0 .  \eqno(2.7) $$
Taking for $q_{min}^2$ the lowest possible Fourier vector $q_{min}^2=(2\pi/L_x)^2$
we obtain 
$$ \gamma_{l.b.} = -\kappa\cdot 4\pi^2/A = -8\pi^2\kappa/N_da  \eqno(2.8)$$
For an individual system $N_d=$const. and $\gamma\sim const./a$; 
otherwise the size dependence at each specific area $a$ is $\sim 1/N_d$.
This is very satisfactory as all these predictions of the theory are
confirmed by our data. The quantity $\kappa$ is the theoretical 
microscopic bending (rigidity) coefficient of the membrane, devoid of any
surface tension in this theoretical picture. 
The scaling (2.8) was obtained independently by Otter[7].

There are other quantities which show a similar kind of discontinuous jump.
Most obviously, the lateral inverse compressibility $K_A$ jumps from 
a high value in region (E) - proportional to $a$ as the slope is common
to all sizes and constant within (E) (see Fig.1) - through a smooth 
transition to a very low value, still positive in the region (C) 
of the floppy bilayer.

The intermolecular energy is another example.  
Fig.3 shows just one plot of $U$ vs. $a$ for the largest system; 
The plot is in accordance with
those of $\gamma(a)$ in Fig.1: two regions of (almost linear) 
variation of $U(a)$  are joined by a transition. 
 The region (C) of negative $\gamma$ corresponds to the region of 
negative slope $dU/da$ here and the region (E) of positive $\gamma$ 
corresponds to the region of positive slope $dU/da$. 
When comparing different sizes, no trend was detected in the
 slopes nor in their difference. The latter
appeared to be independent of the bilayer size. 

 Also histograms of the angle between the axis
of the surfactant dimer and the $z-$axis vary with $a$ with some
(rounded) discontinuity. 
Further examples are found with the parameters of the 
structure factor $S(q)$ in  section III.

In small systems the tensionless state $\gamma(a_0)=0$ belongs to 
the region (E) and the curve $\gamma(a)$ crosses the ordinate axis 
with little change in slope. Also $K_A$ varies little and its value 
at the tensionless state can be used for prediction of $\gamma(a)$ at
a range of values $a>a_0$. It is not so for large systems, where the 
tensionless state falls within the transition region and the derivative
$d\gamma /d a$ changes very fast in the immediate neighbourhood of the 
tensionless state. This can be seen again in Fig.1. 
 
The apparent transition merits investigation.  The derivative of the lateral 
tension w.r.to the 
area, $\gamma^\prime=d\gamma(a)/da$, undergoes a discontinuous jump, 
just like an order parameter would in a first-order transition. This jump is 
apparent from Fig.1 and 2 and can also be seen from plots of the 
computed $\gamma^\prime$ from raw data points.   
 
 However, for a first-order transition the rounding - as resulting from the finite
 size -  should disappear with size increasing indefinitely.
 The area $a_0$ of the tensionless state should stabilize to 
 a definite limit. The point of maximum curvature (maximum change in 
 slope) should merge with the tensionless point $a_0$. 
  
  Fig.4 shows the hypothetical  diagram of $\gamma$ 
vs. $a$ at constant  $T,V,N$ in which rounding disappeared. 
 The plot reduces then 
to two straight lines meeting at a transition point $(a_t,\gamma_t)$.
A straight line $\gamma(a) = +s(a-a_0)$ 
with  positive slope $s$ for $a>a_0$ (region E) is 
continued down to the transition point $\gamma_t,a_t$ with 
 $a_0>a_t$,$\gamma_t<0$. 
 The other line may be a constant or may follow the prediction of eq.(2.8)
as $-\epsilon/a$ with $\epsilon ->0^+$. 
 This plot would be correct for a
 classical first-order transition, sharp in the limit of macroscopic system. 

The derivative $\gamma^{~\prime} $ may be  described by an interpolation formula
$$ f(a)=(f_1+f_2)/2 + (f_2-f_1)\tanh (c~(a-a^\ast))/2~. \eqno(2.9) $$
Here $f_1$, $f_2, c, a^\ast$ are parameters and  
$c$ describes the sharpness  of the transition.  
Integration of $f(a)$ produces a function  for fitting $\gamma(a)$ 
$$ g(a)=g_0 +(f_1+f_2)(a-a^\ast)/2 +(f_2-f_1)\log[2\cosh(c(a-a^\ast)]/(2c) \eqno(2.10) $$
The function $g(a)$ produced  very good 
least-square fits of $\gamma(a)$ for all three sizes with $c$ and $a^\ast$ 
as free parameters. 

However, the sharpness parameter $c$ did not vary significantly with size.
The other parameters varied very little with size either. 
Such patterns are not expected for 
finite-size scaling of first-order transitions for which 
the sharpness parameter increases monotonously with size.

It therefore appears that (2.9) and (2.10) are just 
interpolation formulae for each individual system.

The formulae of Reference[11] which successfully predicted the 
saturation of $\gamma$ to negative values in the floppy regime of 
the bilayer, may also be used for the entire region of 
specific areas. The smoothed mesoscopic area (as 
opposed to theoretical intrinsic area $\bar A$ of the membrane) cannot 
be different from $A\equiv L_xL_y$ as our bilayer is enclosed in a 
box with periodic boundary conditions. With this simplification, but 
without any change of physical meaning of all quantities, we can 
reproduce not too badly the curves $\gamma(a)$ as shown in 
Fig.1. In nondimensional form, appropriate for discussion of size effects,
eq(14) of Reference[11] reads:
$$ u-1+u\sigma/v_2 =(1/c)\log[{{\kappa q_{max}^2 +\sigma}\over{\kappa b u +\sigma}} ] $$
with the abbreviations
$$ u\equiv \bar a/a~~ ;~ c=8\pi\beta\kappa ~~;~ b=8\pi^2/\bar aN_d~.  $$
It is an equation to be solved for $a=a(\sigma)$ or $\sigma =\sigma (a)$.
The parameter $1/v_2 >0$ makes the fixed intrinsic area $\bar a$ slightly elastic.  
There is no doubt that $\sigma$ of [11] is our $\gamma$ as it is coupled to 
$A=L_x^2$.

The relevant parameters have reasonable values: 
$\kappa \sim 5-10 \sim (2.5-5)kT $, $\bar a \equiv 2\bar A/N_d \equiv a_{fixed} \sim 1.13$, 
$q_{max} \sim 2\pi/w $ with the width $w=$4.5, 5, or 6, $q_{min}$ as explained 
above $q_{min}^2=8\pi^2/N_d a$, the elasticity parameter is large 
$v_2\sim 30-40.$. 

However, the agreement is only rough; the transition
predicted theoretically is much too smooth. The excellent fit of the 
floppy region (C) provides reliable values of the bending constant 
$\kappa$ but then a straight line in region (E) is impossible to 
obtain with small values of $\kappa$. Conversely, large values of 
$\kappa$, implying a rigid membrane, can produce a straight line in 
region (E) but then agreement in region (C) is lost. The best overall
reproduction of data was obtained for the smallest system. Also, we
ran into numerical contradictions at the tensionless state; there the 
equations predict a steady shift of $a_0$   to lower values with 
the slope at the tensionless state decreasing to zero and our data 
do not fit this pattern.
 
 Hopefully, the theory sketched in Reference[11] 
  can  perhaps serve as a starting point for improvement  and also 
for  a prediction of the structure factor.

 It is known[12,13] that many 
transitions leave a signature on $C_v$.  
 The heat capacity $C_v$ and the function $C_v(a)$ were also extracted
 from our data. We find a constant $C_v$ within the scatter of data, 
independent of $a$, for all three sizes.

\bigskip
\medskip

\noindent {\bf III. THE STRUCTURE FACTOR AND ITS POLES }.

{\bf A. Extracting and fitting $S(q)$.}

The shape fluctuations of the twodimensional bilayer sheet immersed in 
three dimensions, are similar to capillary waves and it is expected 
that the capillary wave theory may be applicable. However, at the 
tension free state the lateral tension $\gamma$ vanishes. It was 
demonstrated[1], in a simulation experiment,  that the capillary wave
divergence 
$$  S(q) \sim kT/\gamma q^2 \eqno(3.1)$$
is replaced by a stronger divergence
$$  S(q) \sim kT/\kappa q^4 \eqno(3.2)$$
ruled by the rigidity coefficient $\kappa$. This important
result[1] was obtained for the tension-free state $\gamma=0$. In our 
work[5], we have investigated  states with non-zero  $\gamma$ and 
have represented the divergent term as
$$ S(q) \sim 1/(k x^2 + g x)~~~~x\equiv q^2 \eqno(3.3)$$
The structure factor[1,5,7,13,14,15] for the bilayer is related by  
Fourier transformation to the 
correlation $ <h(x,y)h(x',y')>$.

 The function $h(x,y)$ describes the shape of the bilayer by giving
 the "height" or the $z-$coordinate,  as a function of the position 
 in the $x,y$ plane. The latter is the plane of a perfectly flat bilayer.
Now one must define what is meant by the position $h$. 
To avoid a  certain degree of arbitrariness involved in any 
smoothing, we use the actual positions ${\bf r}$ of each head 
of the amphiphilic molecule, as 
$${\bf r}_i =(x_i,y_i,z_i)\equiv (x_i,y_i,h(x_i,y_i)).  $$ 
Therefore in
$$  S({\bf q})= <h_{\bf q}h_{-\bf q}> \eqno(3.4)$$
we take
$$ h_{\bf q} \equiv (1/N_d)\sum_j^{N_d} \exp [i{\bf q R_j}]\times z_j. \eqno(3.5) $$
Here ${\bf R_j}=(x_j,y_j)$ is a two-dimensional position vector 
 of the dimer $j$, $z_j$ its "height", the $z-$coordinate.
 ${\bf q}=(q_x,q_y)$ is the two-dimensional Fourier vector. 
 Care is taken of the translational 
invariance and periodic boundary conditions. 

Unlike here (see also [5,6,7] ), in some simulation work on bilayers the 
$x,y$ grids were introduced together with  a recipe by which 
the local height $h(x_n,y_n)$ would be computed[1,14,15].

The average $<...>$ in (3.4) is  the time average over a
Molecular Dynamics run.

We found earlier[5] and we find now that the divergent term (3.3) {\it must
be} supplemented by terms describing the smooth and regular background.
There are bulk contributions which extend down to $q->0$.

 The structure factor  $S(q)$ determined from the 
simulation was fitted to the semiempirical formula 
$$ S =  1/(k x^2 + g x) +p/x +w_0+w_1 x +w_2 x^2 \eqno(3.6)$$
with $p=0$ or not and with other constants sometimes put to zero.
We introduce the term $p/x$ after Reference[1] where the form 
$1/kx^2 + p/x $ was used. 
Our results for $S(q)$ were obtained along with the results for the 
lateral tension (see preceding Section), in the same Molecular Dynamics runs.
 We show $S(q)$  for nominal sizes $36 \times 36$, $55 \times 55$,
and $100 \times 100$. 
Formula (3.6) represents  very well all
data obtained so far. Formula (3.6) 
is semiempirical in the sense that the first term results from the 
single-mode approximation to the capillary wave hamiltonian[13,16], 
whereas the 
polynomial is there to represent the non-singular bulk-like background
of $S(q)$ in the bulk including the liquid solvent. Also we must not forget
that at or near $q=2\pi/\sigma$ (i.e. near $x\sim 40$)  there appears the 
nearest-neighbour peak. The nearest-neighbour peak is also present
in a distorted form in the height-height correlation functions.
 The rise in $S(q)$ towards that peak begins quite early, 
already near $x\sim 4$ or even less. 
The polynomial in (3.6) takes care of that, within
the limited $q-$range of our data. 

The size dependence of $S(q)$ and the lack of it, are illustrated by 
Fig.5, Fig.6, and Fig.7. These are plots of $S(q)$  against $x\equiv q^2$,
each for the three sizes at equal areas per head $a$. Fig.5 is for such 
$a$'s that $\gamma\sim 0.84$; the systems are clearly inside the region (E),
as defined in  Section II. Fig.6 is for the tensionless states, or almost
tensionless, and Fig.7 is for small $a$, negative $\gamma$, thus deep 
inside region (C) of floppy  bilayers.

These three Figures demonstrate the behaviour of the singular term (3.3).
As reported below, $k$ is always positive and $k>1$. Then for positive 
$g$ as $x\rightarrow 0$ the curve flattens a little as it crosses over 
from $(kx^2+gx)^{-1}$ to $1/gx$ (see Fig.5). 
For $g$ nearly vanishing (see Fig.6) this flattening disappears and 
the rigidity divergence $1/kx^2$ dominates. For $k>0, g<0$ (see Fig.7) 
there is a pole on the real axis at
$$  x = x^\dagger \equiv (-g/k)~. \eqno(3.7)$$
Then  $S(x)$ diverges faster than it did, aiming at
 infinity at the asymptote $x=x^\dagger$.
 This is what we see in Fig.7
where each system, depending on its size, has its own asymptote.
Otherwise at higher values of $x$, the size appears to affect $S(q)$ 
very little. 
The poles of (3.3) and (3.6) are discussed further below in subsection B.

The  data on $S(x)$ were fitted to (3.6). 
Since one has to be careful with  least-square 
fits, as a precaution we have used several versions of (3.6) 
for each set of data, 
(a) with $p\neq 0$ and with $p\equiv 0$, 
(b) with $w_2$ either zero or not 
(c) with $w_1$ zero or not.

Fits proved to be robust in most cases but gave scattered 
results for the coefficients $k$ and $p$ for systems with large
$\gamma$ and area $a$. One firm conclusion is that {\it always} 
the coefficient $k$ was positive $k>0$; not a single instance was 
ever found with $k=0$ or $k<0$.
Thus $k$ {\it can} be interpreted as a rigidity coefficient (up to a constant).

Parameter $p$ was sometimes erratic but other parameters, representing 
in (3.6)the smooth background, did vary smoothly with $a$ and not much.
Protrusions, accounted for as $p/x$, are small-wavelength fluctuations 
and physically
a protrusion may be hardly distinguishable from a nearest-neighbour 
interaction. That may explain a correlation of the least-square $p$ with 
$w_i$'s.

Fig 8. shows the variation of the least-square parameter $k$
with the area $a$. Referring to the two regions (C) and (E) described
in section II, we can distinguish the (E) region of fast increase of 
$k$ with $a$, so that the stiff bilayer of high $\gamma$ at large $a$
has a high rigidity, and the buckling  floppy bilayer
at low $a$ and negative $\gamma$ in region (C) has a relatively low $k$.
This seems plausible. In region (C) the low $k\sim 5.$ stays constant 
whereas in region (E) it increases overall, in some fits linearly,
in some very little.  
Eliminating $a$ we obtain the 
  variation of  least-square parameters with 
$\gamma$; Fig.9 shows the overall increase of $k(\gamma)$. 
 
Fig.10 shows how closely the least-square coefficient $g$ follows $\gamma$. 
Nevertheless consistently $g > \gamma$, suggesting these are two different 
quantities.
In fact, it may be argued that $g$ describing the spontaneous 
distortion-fluctuation of the interface, is coupled to the hypothetical 
"true area", whereas $\gamma$ is coupled to the "projected area" fixed by
the experimentalist. This issue was discussed also recently[7]. 
These considerations go back to the distinctions between 
different molecular expressions[9] for the  surface tension of the 
liquid interface. It appears[17] that in a "normal" interface 
$g\equiv \gamma$ and  in a membrane-like sheet of a bilayer, 
$g$ from $S(q)$ need not be equal to $\gamma$. Consistently, as the 
Figure shows, $g > \gamma$.

{\bf B. The presence of a pole on the $q-$axis.}

Returning now to Fig.7 which shows $S(q)$ for 
negative $\gamma$ we notice that increasing the size essentially increases
the range as the asymptote at $x=x^\dagger$ is pushed to the 
left to lower values. All the divergences are well handled by (3.3) and (3.6),
but for $g<0,k>0$ the $x-$axis is divided into two parts $0<x<x^\dagger$ and 
$x^\dagger < x$. For $x>x^\dagger$ all is well, but for $0<x<x^\dagger$ 
the expressions for $S(q)$ become negative and $S\rightarrow -\infty$ as 
$x$ tends to $x^{\dagger~}$ from below. 
Clearly in the interval $0<x<x^\dagger$ the expression for $S$ is 
unphysical $-$ as the very definition of the scattering factor $S$ makes it
a positive  quantity. The interval  $0<x<x^\dagger$ must be excluded.
As a consequence, we cannot consider the limit $q\rightarrow 0^+$ or $x\rightarrow 0^+$ , as is 
usual for a discussion of the capillary waves and of the "capillary 
divergence". 

The resolution of these points comes from the realization 
that in defining $S(q)$ as the Fourier transform of a correlation 
function in the ${\bf r}-$space, in fact we are dealing with a 
Fourier {\it series}. Indeed  the positions ${\bf r}$ vary continuously 
but the ${\bf q}-$vector, $(q_x,q_y)$, is restricted to 
$$  q_x = ({{2\pi}\over{L_x}}) n_x~~~~  (n_x=0,\pm 1,\pm 2,...) \eqno(3.8)$$
and similarly for $q_y$. The case $n_x=n_y=0$ being excluded, 
the lowest possible value of $\vert q \vert$ is  
$q_0=2\pi/L_x$ and $x$ cannot be smaller than 
$$ x_0 \equiv 4\pi^2/L_xL_y. \eqno(3.9)$$
Thus the size of the system is ever present in $S(q)$. For positive 
$k$ and $g$ one can speak of the limit $q\rightarrow 0$, but 
for negative $g$ this is not correct. 

Knowing that for given size $x$ must fulfill $x>x_0$ we must check 
that the asymptote $x=x^\dagger$  is always beyond reach, i.e. that
$$ 0<x^\dagger < x_0=4\pi^2/A \eqno(3.10)$$ 
is always fulfilled. This is the case with all our data and with all 
our least-square fits. Never any least-square fit had to be rejected 
because the pole would fall in the physical region $x_0<x$.

Now we consider the size dependence of the position of the pole.
 As the parameters $k$ and $g$ vary smoothly (allowing 
for scatter) so does $x^\dagger$ and therefore  the 
calculated values of $(-g/k)$ for some small positive $g$ are also 
included. 

Of great significance is the plot in Fig.11 
of $x^\dagger$ against $\gamma$ because 
here the different sizes fall on a common curve which, with reasonable 
accuracy, aims at the point (0,0). This is very satisfactory: it means 
that with the increased system size the 
position of the asymptote $x=x^\dagger$  tends to $0^+$. 
Thus the unphysical region
$0<x<x^\dagger$ shrinks to zero with the increased system size.

 Such behaviour  appears very satisfactory and also in agreement 
with the content  of Section II. 

In this way we have resolved a serious difficulty which was either 
ignored or misinterpreted in the past.

\bigskip
\medskip
\noindent {\bf IV. SUMMARY and DISCUSSION.  }

We used atomistic simulations of dimer molecules forming reverse bilayers
in a  solvent. We obtained results on the bilayer isotherm $\gamma(a)$,
the apparent buckling transition, the scaling the negative lateral tension if 
the floppy bilayer, the size dependence of the structure factor $S(q)$, and on 
the new divergence of $S$.

 These results emphasize again the
 deep differences between the membrane-like two-dimensional sheets of 
 surfactants and the transition region between two coexisting fluid 
 phases that is given the name of an interface. 

 The bilayer peculiarities were studied in Section II; the lateral 
 tension of a stable bilayer though given by (2.1) can be null or negative. 
 This is possible owing to the
 existence of a proper or intrinsic area $\bar A$ of the membrane, constant 
 in a first approximation.
 In addition to the thermodynamic parameters, the state of a bilayer 
 depends on the externally enforced area - the edge of the box in 
the simulation. The function $\gamma(a)$ is the the bilayer {\it isotherm}. 

The few simulations that 
have gone beyond the tensionless state and have determined the 
bilayer isotherm for a large range of areas[2,4,5,6,7], all found 
interesting and nontrivial behaviour.  We find size-independent 
linear $\gamma=26.(a-a^\ast)$ for large $\gamma>0$. In the Lennard-Jones
units (Appendix A) the slope is $26.\epsilon /\sigma^4$. 
 In small systems[1,2,5] the isotherm continues down 
with little change in slope, but in bigger systems
there is a very quick change of slope and a transition to a floppy 
state of a buckling bilayer with  negative $\gamma$  of a flat 
isotherm saturating to the size dependence (2.8). The constraint of 
constant intrinsic area makes the bilayer to buckle. 
Negative $\gamma$'s can be extremely close to zero. 
 
  The bilayer together with the solvent fluctuates owing to thermal motion.
In larger systems the undulations destroy the extension of the linear portion
of the isotherm into negative $\gamma$'s; the tensionless state lies in the 
region of fast change  of the slope. It is an unwelcome piece of 
news. It means that the inverse compressiblity $K_A$ taken at the 
tensionless state, not only is not easily  determined accurately 
but moreover is not representative - unless the system is {\it small} enough.
 Rather, the slope of the linear part, common to different sizes,
 is representative.

We also find that that changeover, from region (E) 
of the common slope to region (C) of negative $\gamma$, is 
very sudden and abrupt, suggestive of a buckling transition[18]. However,
such a transition, in the strict meaning of a mathematical singularity,
has not been predicted for a bilayer. We  find 
no definite increase in sharpness with size.  
In bilayers made of long-chain surfactant molecules[6,7],
the transition  of smaller systems was sudden and appeared 
truly discontinuous; for large bilayers
the shape of $\gamma(a)$ became the same as in Fig.1. A picture emerges 
of a discontinuous transition which is destroyed by the 
increase in size of the  bilayer.
 
The structure factor $S(q)\sim <h_qh_{-q}>$ was independent of size 
as long as $a>a_0$; in the region of floppy bilayer the strong size 
dependence appears and  the new pole at $q>0$  replaces the capillary
wave divergence  at $q->0^+$. This issue is now satisfactorily resolved
(see Section IIIB). 

The theoretical picture leading to good predictions[11] for the floppy
bilayer, is very attractive in its simplicity: 
as the area $\bar A$ is constant, 
there is no surface tension contribution, only bending. 
Free energy expressions i.e. 
 mesoscopic hamiltonians are constructed building on such ideas[11].
   However it is only but a first step; the predicted transition is too smooth, 
not all size dependence is correctly obtained (see Section II), 
and the structure factor $S(q)$ is not known; it ought to appear  
{\it with another coefficient} larger than $\sigma$ (see 
Fig.10).

 Finally, we remark that the relations between $\kappa$ (i.e. $k$) and 
 the inverse compressibility $K_A$, taken from the theory of elasticity
of plates[1], seem to do well in bilayers made of long chains[1,7] but 
fail for dimers studied here. 

\bigskip
\medskip
\noindent {\bf V. ACKNOWLEDGEMENTS  }.

This work was financially supported in part by the 
Komitet Badan Naukowych grant Number 4T090A-05025. The author 
is indebted to Docent A. P. Poniewierski and Professor R. Holyst for 
early discussions. 

\vfill\eject
\bigskip
\medskip
\noindent {\bf VI. APPENDIX A  }.

The units are those commonly used for systems with Lennard-Jones (6-12)
potential 
$$ u_{00}(r)= 4\epsilon((\sigma/r)^{12} -(\sigma/r)^{6}) \eqno(A.1)$$
The unit of energy 
 is $\epsilon$ and the unit of length is $\sigma$.  Areas $A$ or $a$
are given in units of $\sigma^2$, free energy and energy in units of 
$\epsilon$, reduced temperature $T^* \equiv kT/\epsilon$ where $k$ is the 
Boltzmann constant, reduced lateral tension 
$\gamma^*=\gamma\sigma^2/\epsilon$, reduced pressure 
$p^*=p\sigma^3/\epsilon$, compressibility 
modulus $K_A$ in units of $\epsilon/\sigma^2$. In the text all quantities are reduced quantities 
and the asterisk is dropped from the notation.  

In this work all constituent particles are spherical and interact according
to (A.1). The system contains $N_a$ "a" solvent particles and $N_d$ surfactant
dimers made of $N_d$ further "a" particles and of $N_d$ "b" particles. The 
constituents of the surfactant "a-b" dimer are bound by an unbreakable 
but flexible chemical bond. 
 Particles are of equal sizes and have equal masses. The potentials are
cut and shifted[19,20]; for a pair $\alpha - \beta$ 
$$ u_{\alpha\beta}(r) = (u_{00}(r) -u_{00}(r^c_{\alpha\beta}))
\times \eta(r^c_{\alpha\beta}-r) \eqno (A.2)$$
where $\alpha$=a,b and  $\beta$=a,b.
The Heaviside function is $\eta(x)=1$ or 0 if $x>0$ or $x<0$.   
For like particles, $\alpha=\beta$, the cutoff distance  is taken at
$$r^c_{\alpha\beta}=2.5~~. \eqno(A.3)$$ 
(in units of $\sigma$). For $\alpha \neq\beta$, i.e. for "a,b" pairs  
$$r^c_{\alpha\beta}=r^*\equiv 2^{1/6}\sigma \eqno(A.4)$$
makes the potential purely repulsive in the spirit of WCA[3,5,6,19].

The bilayers studied in this work are "reverse" bilayers so named 
in analogy to the reverse micelles. The reverse bilayers were constructed
and found to be stable[5]; are formed not by entropic effects in conjunction
with the hydrophobic effect, but  primarily  by energetic preference. The 
latter is achieved by making the "b-b" attraction much stronger.
To achieve this, we modify the energy depth parameter $\epsilon$ for the 
pairs "b-b" (only), keeping the common value of $\epsilon$ for all other 
pairs. That is  
$$ \epsilon_{bb}=S\epsilon_{aa}.  \eqno(A.5)$$
where the strength parameter $S$ is to be
choosen  $S>>1$. Then the "b"-particles, the "b"-ends of the 
surfactant dimers, make the strongly cohesive core of the bilayer
whereas the weakly attracting "a"-ends stay outside in contact with 
the "a"-solvent. 

  The Molecular Dynamics simulations were done in a well-established manner: 
at constant volume $V$, and particle numbers $N,N_d$, by using Verlet 
leap-frog algorithm and Nose-Hoover thermostat[19,20]. In all series quoted 
here the reduced temperature was $T=1.9$ and the overall density 
was a reasonably high liquid-like density of $\rho=N/V=0.89204$.
The augmented "b-b" strength parameter was $S=4.$
The total number of particles $N$ contained $N_d$ dimers that is 
$N_d$ "b" particles all of them bound in "a-b" dimers, $N_d$ "a"
particles also permanently bound in  "a-b" dimers, and $N-2N_d$ 
free "a" particles which were the solvent.
Given the numbers $N,N_d$ and the above density $\rho$ the volume $V$
was fixed; once the area $a$ was chosen, $A=N_d a$ followed, hence
$L_x=L_y$ as the square root, whereas $L_z$ was adjusted to fit the 
volume $V=A L_z$. The areas $a$, as shown in Fig.1 were chosen to be
in the interval of stable bilayers.

\noindent For the size (iii) (nominally of area $36 \times 36$) \hfill\break
$N=40000,N_d=2238,N-2N_d=35524, 33.<L_x<35.5, 41.2<L_z<36.6$

\noindent For the size (ii) (nominally of area $55 \times 55$) \hfill\break
$N=160000,N_d=5760,N-2N_d=148480, 51.<L_x<57.8, 69.<L_z<53.5$

\noindent For the size (i) (nominally of area $100 \times 100$) \hfill\break
$N=281344,N_d=18720,N-2N_d=243904, 93.<L_x<105., 36.5<L_z<30.$

The proper normalization of the structure factor $S(q)$ leads to the
prediction $<h_qh_{-q}>\equiv \hat S(q) = (kT/L_xL_y)\cdot {{1}\over{kx^2 +g x}}$
for the singular part of $S$.
The  size-independent quantity is $ A\hat S$. The Fourier components  of 
$h(x,y)$ were calculated  for each monolayer separately according 
to (3.5); the number of molecules in either will vary with time.
    
\vfill\eject
\medskip
\noindent {\bf VII. REFERENCES  }.
\medskip

 \item {[1]}  G. Gompper, R. Goetz, and R. Lipowsky, Phys. Rev. Lett. {\bf 82}, 221 (1999);
 \item {[2]}  R. Goetz and R. Lipowsky, J. Chem. Phys. {\bf 108}, 7397 (1998).
 \item {[3]}  B. Smit, Phys. Rev. A {\bf 37}, 3431 (1988).
 \item {[4]}  O. Farago and P.Pincus, J. Chem. Phys. {\bf 120}, 2954 (2004);
 for definition of the model see O. Farago, {\it ibid.} {\bf 119}, 596 (2003).
 \item {[5]}  J. Stecki, J. Chem. Phys. {\bf 120}, 3508 (2004).
 \item {[6]}  J. Stecki, J. Chem. Phys. Comm.{\bf 122}, 111102 (2005).
 \item {[7]}  W. K. den Otter, J. Chem. Phys.{\bf 123}, 214906 (2005).

\item {[8]}  J. G. Kirkwood and F. P. Buff, J. Chem. Phys. {\bf 17}, 338 (1949);
 since, the formula has been used and rederived many times; for an application
 to bilayers with flexible chains, see [2]; for interface elasticity, see
 J. Stecki and P. Padilla,  J. Chem. Phys. {\bf 104}, 7249 (1996); for an
 extension to higher orders see J. Stecki, Mol. Phys.{\bf 100}, 2555 (2002).
There is not a slightest doubt that it is the correct recipe for computing 
the tension coupled to the edge of a parallelepipedian box, $L_xL_y$.

 \item {[9]}  J.~S. Rowlinson and B. Widom, 
{\it Molecular Theory of Capillarity} (Clarendon, Oxford, 1982). For a 
derivation of the K.-B. formula see page 89. 
   
 \item {[10]} simulations of model bilayers were mostly done for model 
long-chain molecules and models of lipids,  
producing or attempting to produce the tensionless state. For a rare
example of a $\gamma(a)$ dependence see  S. E. Feller and R. W. Pastor,
  J. Chem. Phys. {\bf  111}, 1281 (1999), {\it ibid.} {\bf  103}, 10267 (1995);
[2], reference [14], [15]  below. 

\item {[11]} J.-B. Fournier, A. Adjari, and L. Peliti, Phys. Rev. Lett. {\bf 86}, 4970 (2001),
             where also references to older theoretic considerations can be found.

\item {[12]} D. Nelson, in D. Nelson, T. Piran, and S. Weinberg, Eds., {\it Statistical
 Mechanics of membranes and Surfaces} (World Scientific, Singapore, 1989).

 \item {[13]} P. M. Chaikin and  T. Lubensky, {\it Principles of Condensed Matter 
                Physics} (Cambridge University Press,1995) page 628ff.
 \item {[14]} E. Lindahl and O. Edholm, Biophys. J. {\bf 79},423 (2000). 

 \item {[15]} S. J. Marrink and A. E. Mark, J. Phys. Chem. B {\bf 105}, 6122 (2001).

 \item {[16]} see e.g. J. Stecki, J. Chem. Phys. {\bf 114}, 7574 (2001); ibid.
 {\bf 108}, 3788 (1998).
 \item {[17]}  Our old liquid-vapor simulation data[12] produced 
 the surface tension from the Zwanzig-Triezenberg formula[9] equal to within
 0.01percent to $\gamma$ from the virial (i.e. 
 Kirkwood-Buff) expression. Here the constraint of constant or almost constant
 membrane area prevents this equality.  How the 
 constraint modifies the capillary waves is not known[11], but $g>\gamma$ 
systematically (see Section II).

\item {[18]} The established terms such as "Kosterlitz-Thouless transition",
"melting transition", ...,  indicate  a mathematical 
singularity or discontinuity, whereas, on the other hand, "transition to ..."
such as "transition to buckled state", can refer to a smooth change.
For those reasons we avoid the syntax "buckling transition" because a
true singularity, especially in large systems, appears 
elusive. 

\item {[19]} S. Toxvaerd (University of Copenhagen), private communication; 
see also
  L. C. Akkermans, S. Toxvaerd and W. J. Briels, J. Chem. Phys. 109, 2929 (1998).

 \item {[20]} see e.g. S. Toxvaerd,  Mol. Phys. {\bf 72}, 159 (1991); 
Phys. Rev. E {\bf 47}, 343 (1993). 

\vfill\eject
\medskip
\noindent {\bf VIII. FIGURE CAPTIONS    }.
\bigskip
Caption to Fig.1

The lateral tension $\gamma$  plotted against area per dimer head, $a$, for
three sizes : size (i) with diamonds, size (ii) with plus signs, size (iii) 
with squares. All points are at $T=1.9$, high liquid-like density 
$\rho=0.89204$;
size (i) is that of an area about $100. \times 100.$, size (ii) - of an area 
about $55.\times 55.$, and size (iii) - of an area about 36. by 36.. 
See the Appendix  for further details. $A$ is total area, $N_d$ is the 
number of dimer heads, $a$ area per head. Note common curve for large $\gamma>0$.

\bigskip
Caption to Fig.2

The plateau of $\gamma(a)<0 $ at small $a$, plotted against 
$1/A \equiv L_x^{-2}$ is shown to follow the line $y=300*x$. See text.  
The near-constant plateau tends to zero with increased system size. Logarithmic 
plots confirm this Figure. For system parameters see  Appendix A.

\bigskip
Caption to Fig.3

Total intermolecular energy per particle, $U$,
plotted against $a$, area per dimer head, for the largest system of nominal 
size $100 \times 100$. Note a discontinuity, though rounded, in the 
derivative.

\bigskip
Caption to  Fig.4    

The $\gamma(a)$ plot for a hypothetical first-order transition in a 
virtually infinite system:
fall in $\gamma$ with decreasing $a$ towards $a_0$ in region (E);
 a discontinuous change of slope to a near-vanishing value 
in region (C).  See text. 

\bigskip
Caption to  Fig.5

The structure factor $S(q)$ plotted against $x\equiv q^2$ for three
system sizes at the same area per head $a=1.09$ :
Diamonds - size (i) $100 \times 100$; plus signs - size (ii) $55 \times 55$;
square boxes - size (iii) $36 \times 36$. The line shows the fit of the
function (3.6) to data on system of size (i). The size dependence is 
seen to be negligible, except for the extension of range of $q^2$. The 
lateral tensions are strongly positive $\gamma = 0.85-0.90$.

\bigskip

Caption to Fig.6

The structure factor $S(q)$ plotted against $x\equiv q^2$ for three
system sizes at the same area per head $a=1.05$ :
Diamonds - size (i) $100 \times 100$; plus signs - size (ii) $55 \times 55$;
square boxes - size (iii) $36 \times 36$. The line shows the fit of the
function (3.6) to data on system of size (i). The size dependence is 
seen to be negligible, except for the extension of range of $q^2$. The 
lateral tensions are near zero: $\gamma$ = 0.02-0.05. Compare with preceding
and following Figure.

\bigskip
Caption to   Fig.7 

The structure factor $S(q)$ plotted against $x\equiv q^2$ for three
system sizes at negative $\gamma$ and in a floppy state
( area per head $a \sim 0.973-0.993$).
Diamonds - size (i) $100\times 100$; crosses - size (ii) $55 \times 55$;
square boxes - size (iii) $36\times 36$. The thick line shows the fit of the
function (3.6) to data on system of size (i); the two other lines
are there to guide the eye. The size dependence is 
seen to be negligible, except for the extension of range of $q^2$. The 
lateral tensions are all negative but in the crupling floppy state (C),
 depend now on size. Compare with preceding two Figures.

\bigskip
Caption to  Fig.8

The variation of the least-square parameter $k$
with the area $a$. Note $k>0$. Diamonds - size(i) $L_x\sim 100$,
Crosses - (ii), 55, 
Squares - (iii), 36.
Note  the two regions (E) and (C); (C) buckling floppy bilayer with 
low and constant $k$ and (E) extended tensioned bilayer with $k$ 
roughly linear with area per head $a$.

\bigskip
\vfill\eject
Caption to          Fig.9

The variation of the least-square parameter $k$
with the lateral tension $\gamma$. Diamonds -  size(i) $L_x\sim 100$,   
Crosses - (ii), 55, 
Squares - (iii), 36.
Note overall increase of $k$ with $\gamma$.

\bigskip
Caption to          Fig.10	   	 

The variation of the least-square parameter $g$
with the lateral tension $\gamma$. Diamonds - size(i) $L_x\sim 100$, 
Crosses - (ii), 55, 
Squares - (iii), 36.
Note how closely $g$ follows $\gamma$ though systematically $g$ is the 
larger of the two.

 \bigskip
 Caption to          Fig.11

Pole position $x^\dagger$ plotted against the lateral tension 
Diamonds -	 size(i) $L_x\sim 100$,   
Crosses - (ii), 55, 
Squares - (iii), 36. See text.

end of Captions. end of paper.

\vfill\eject
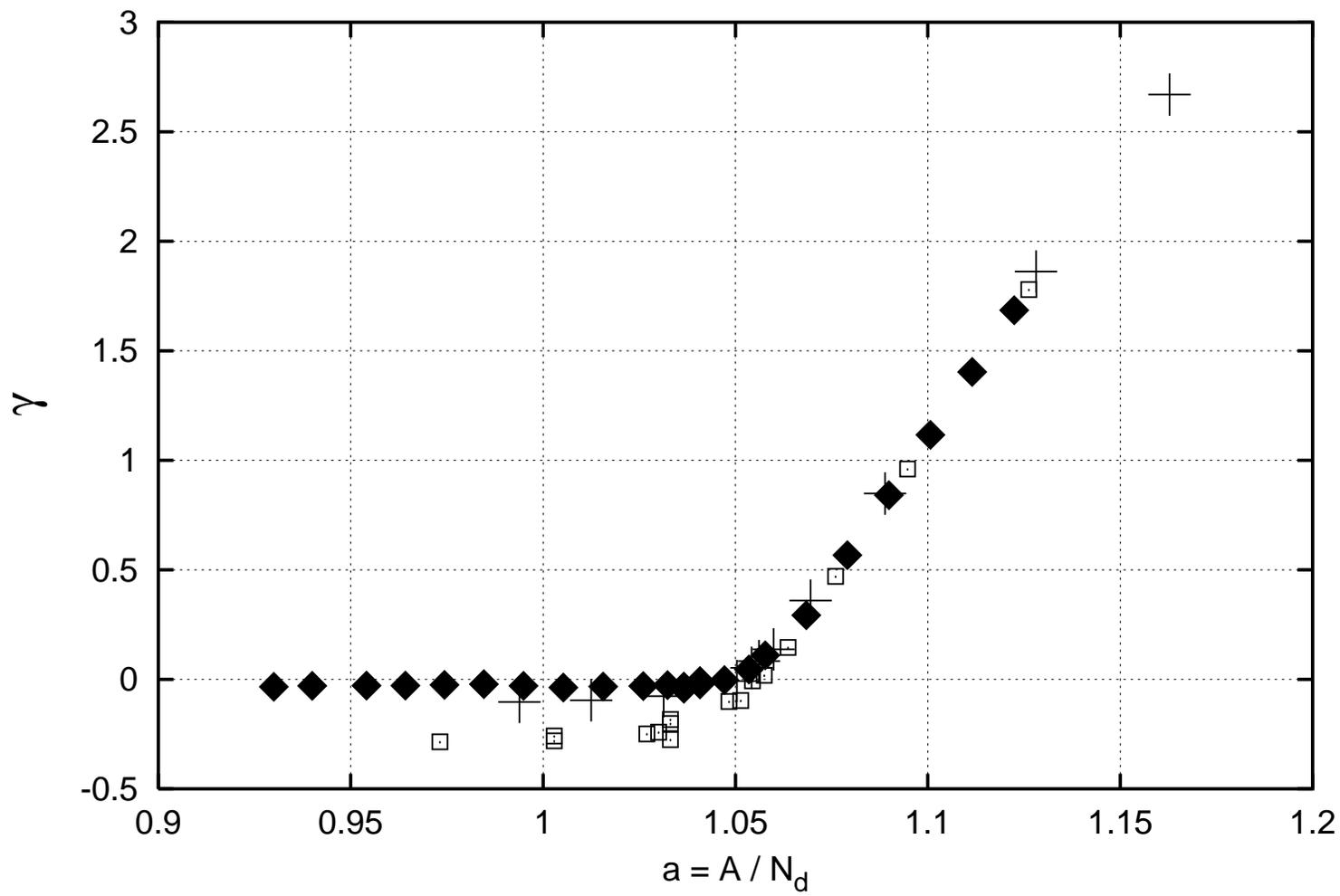
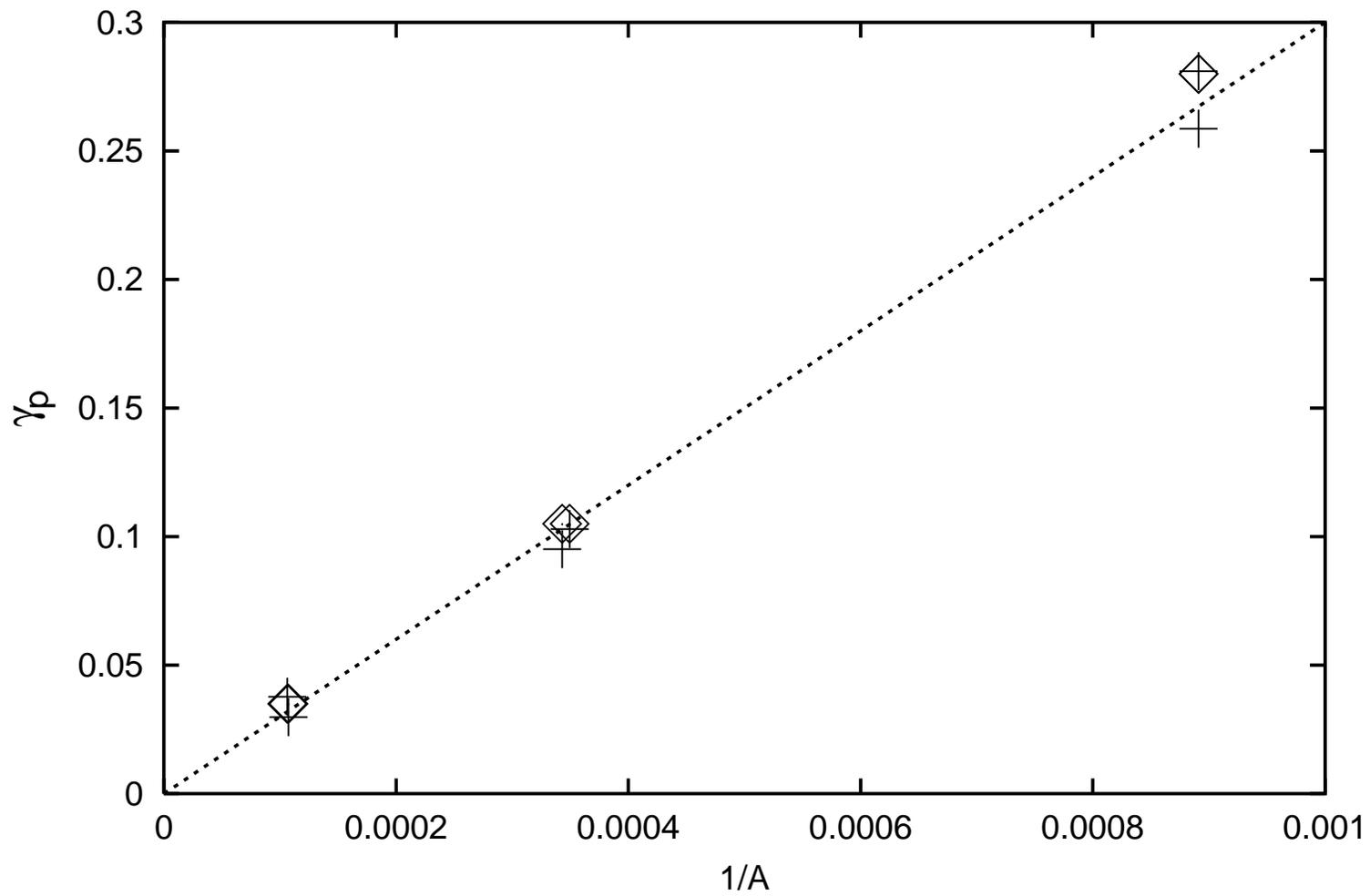
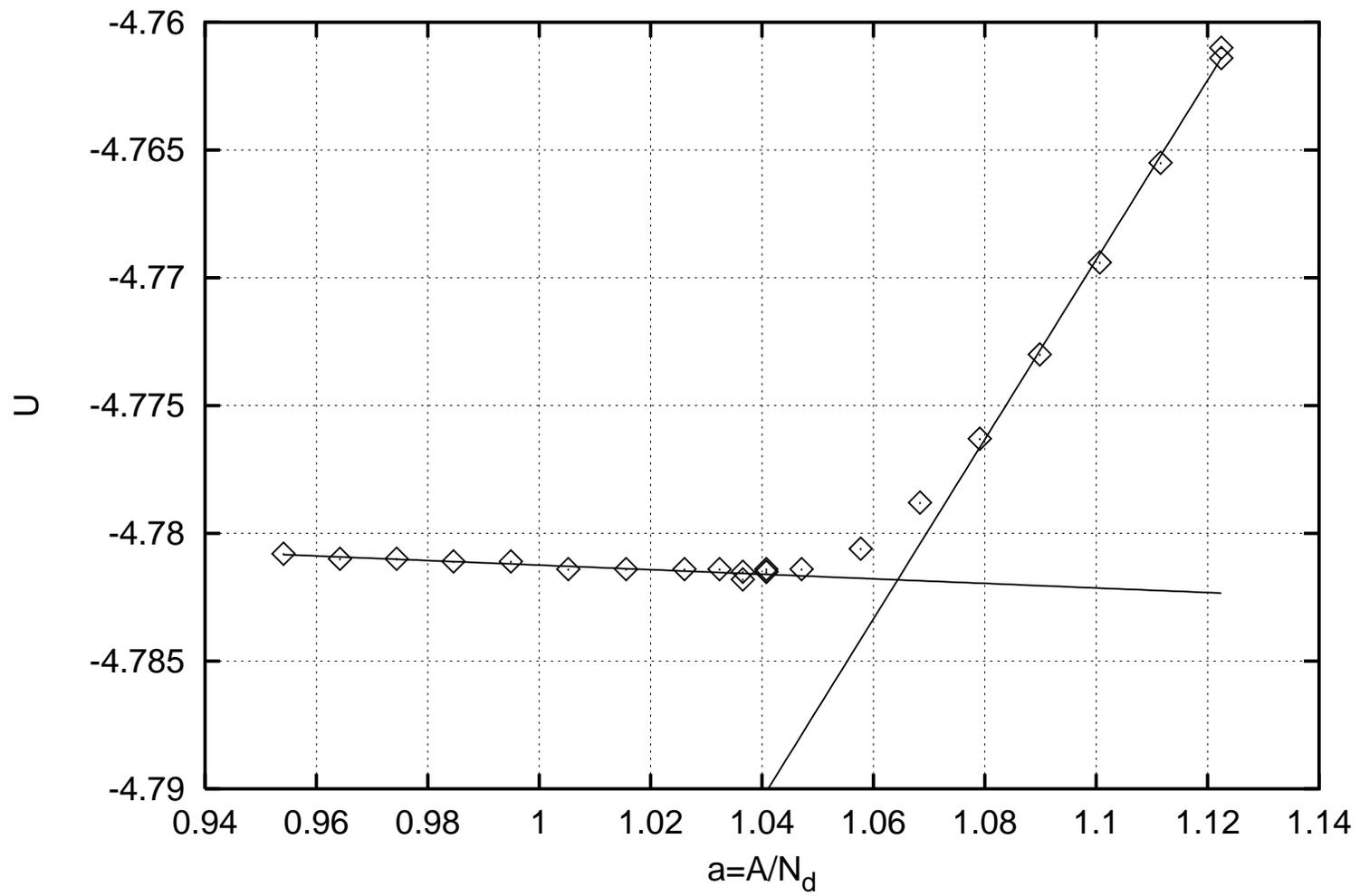
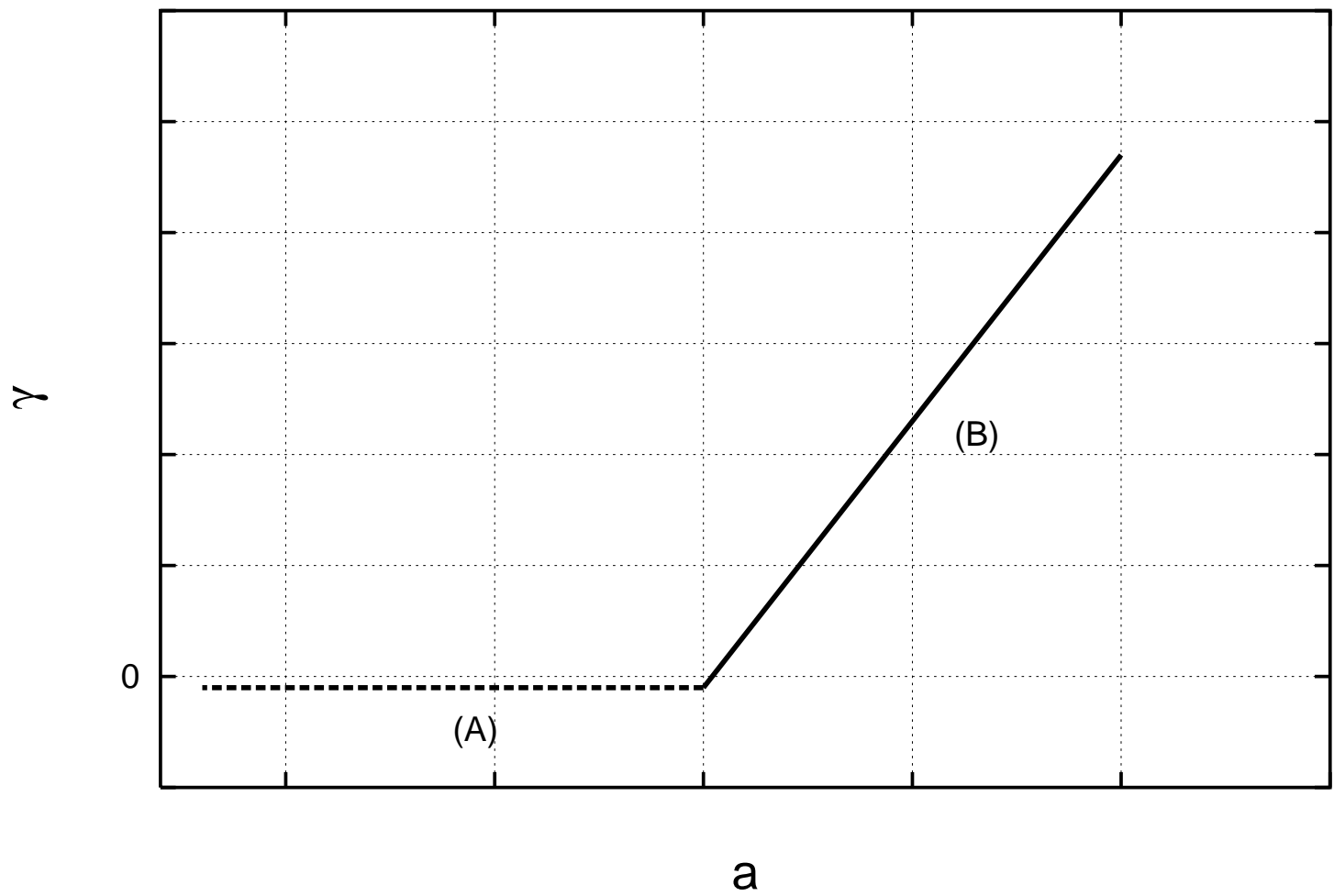
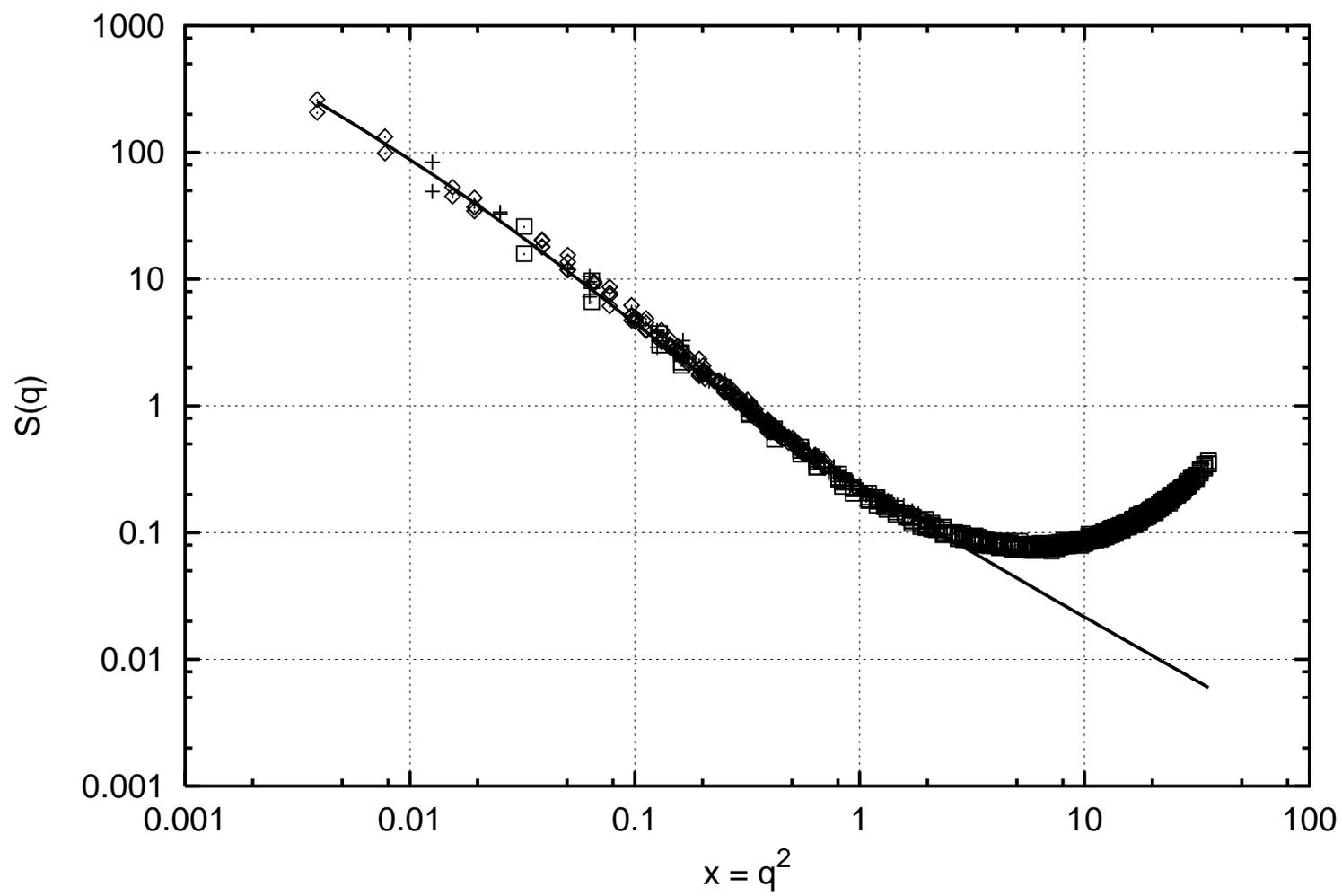
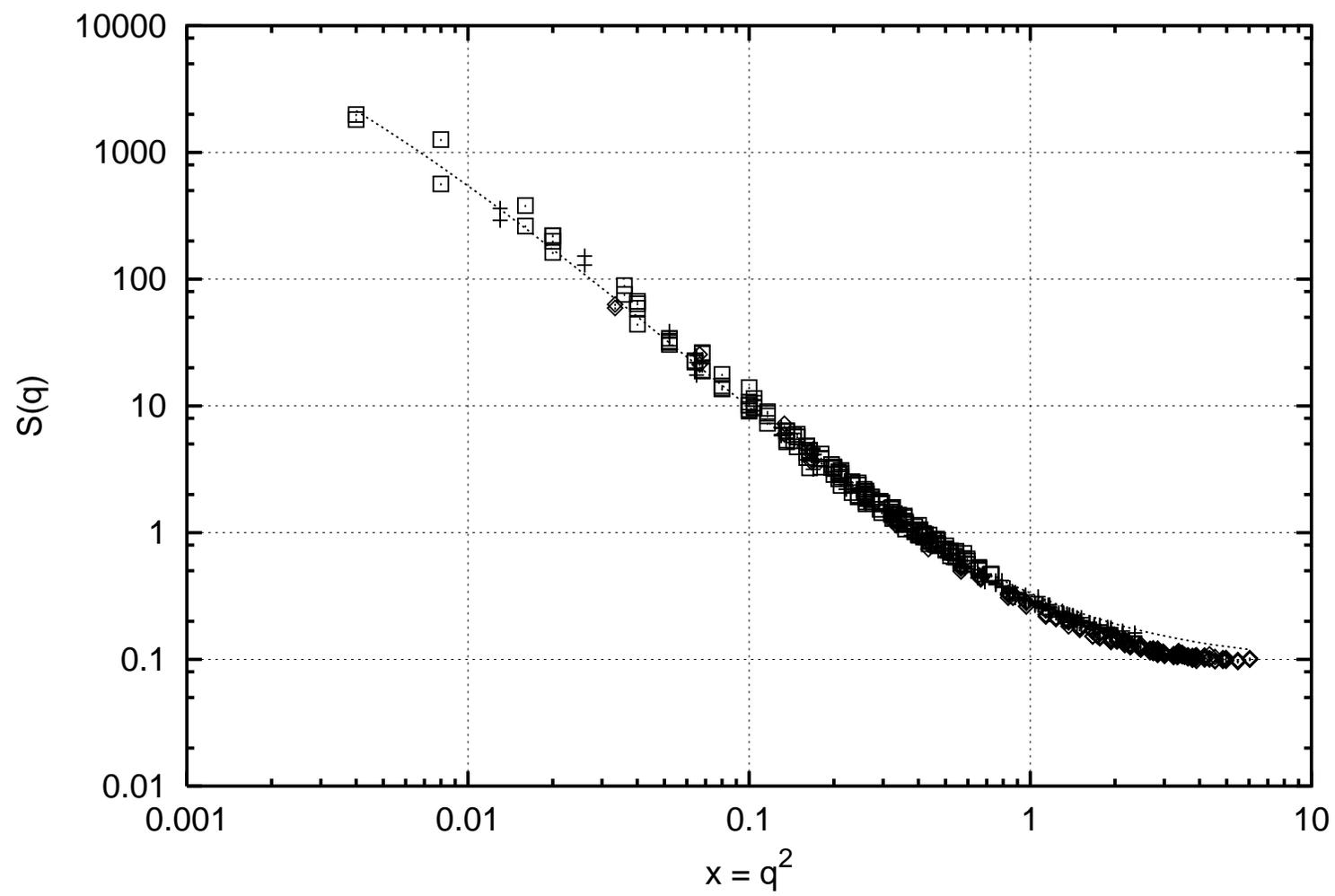
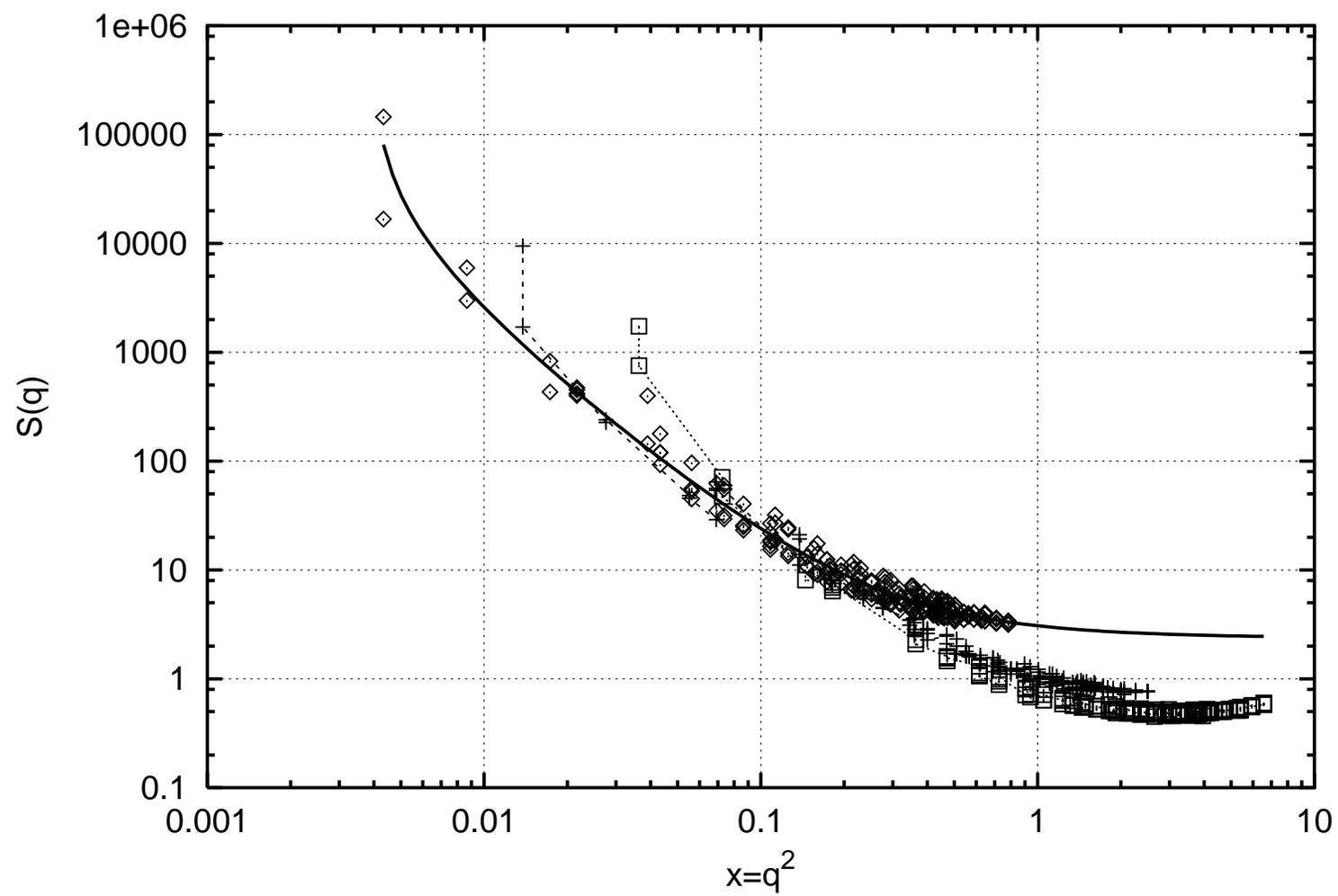
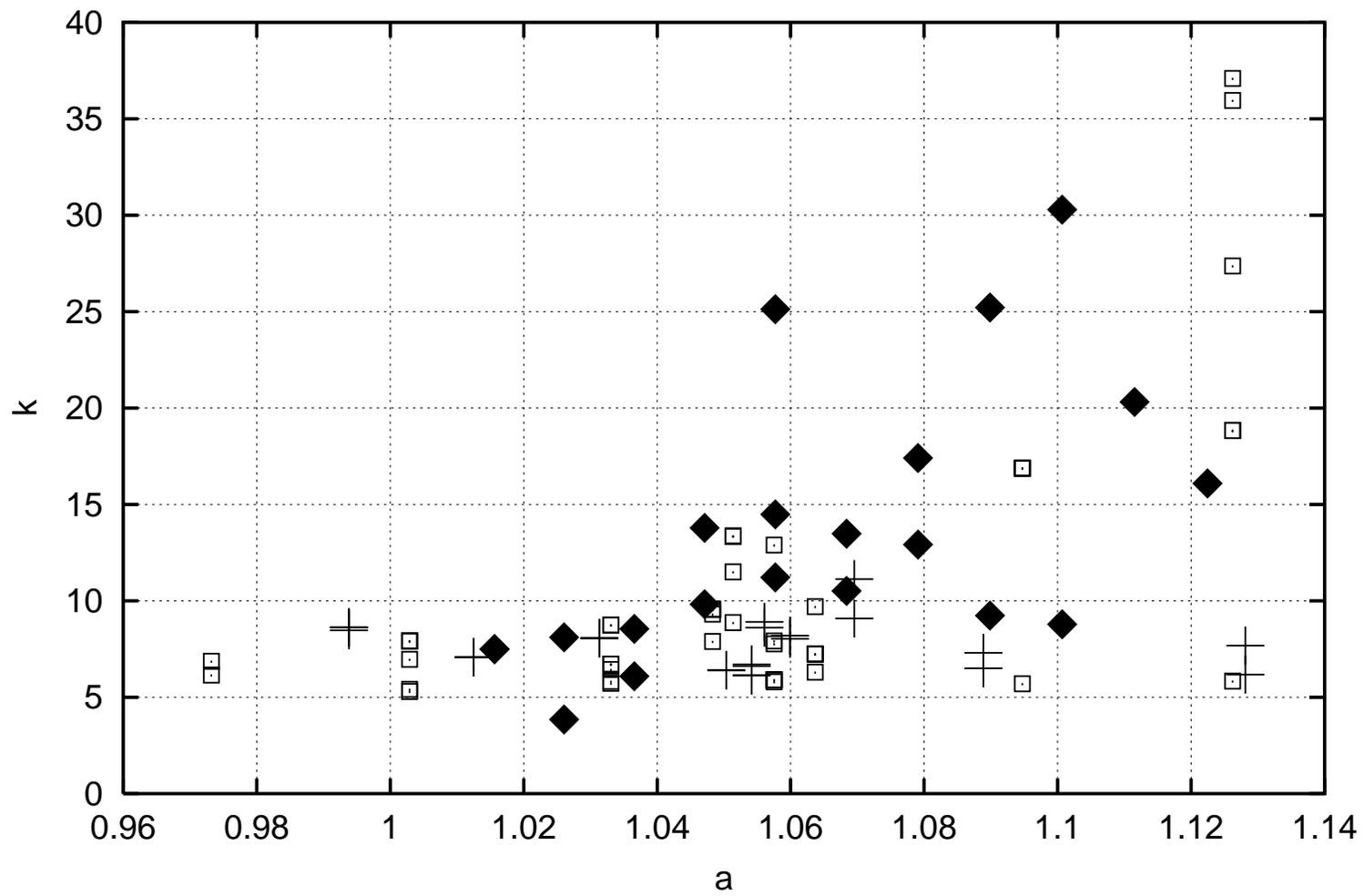
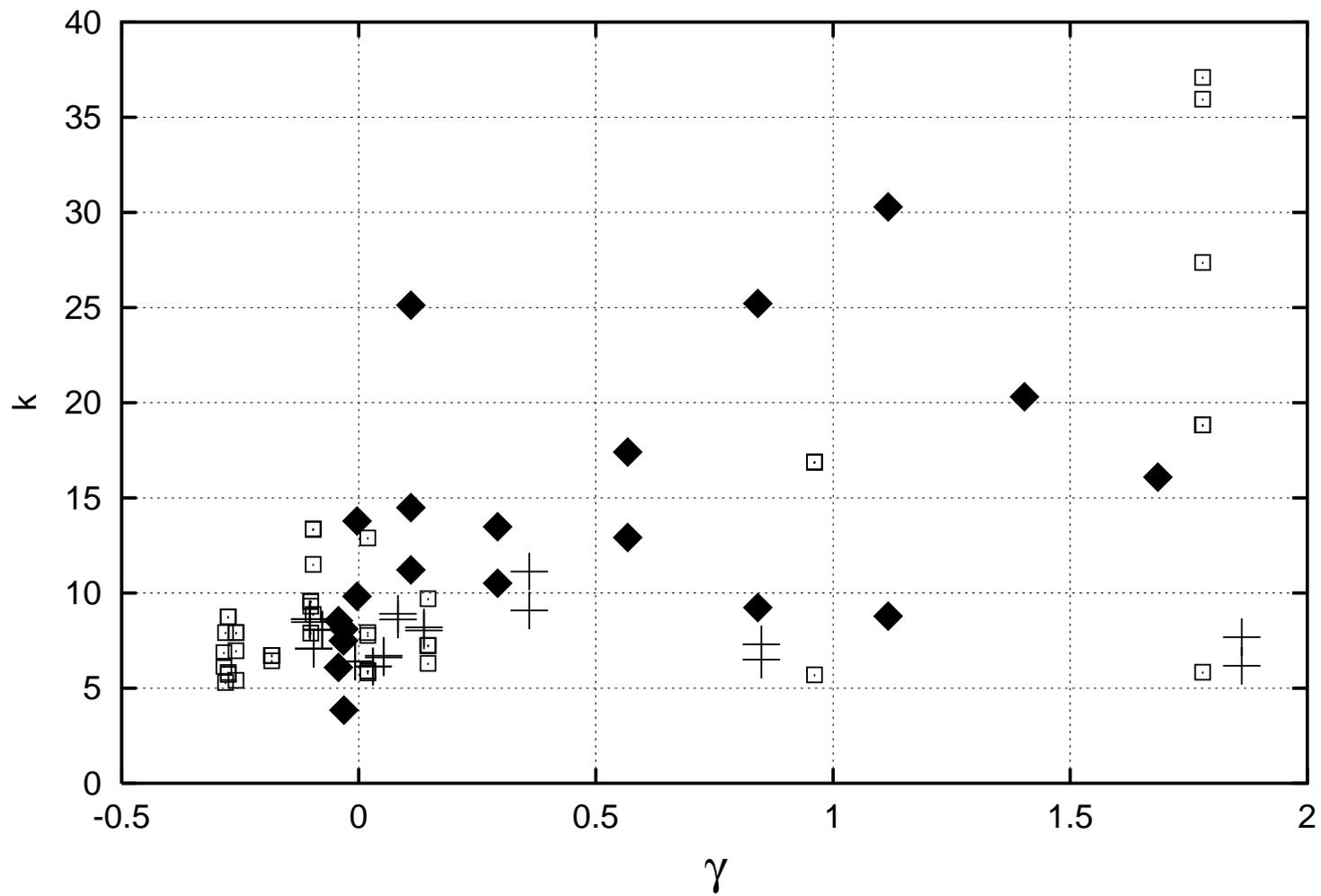
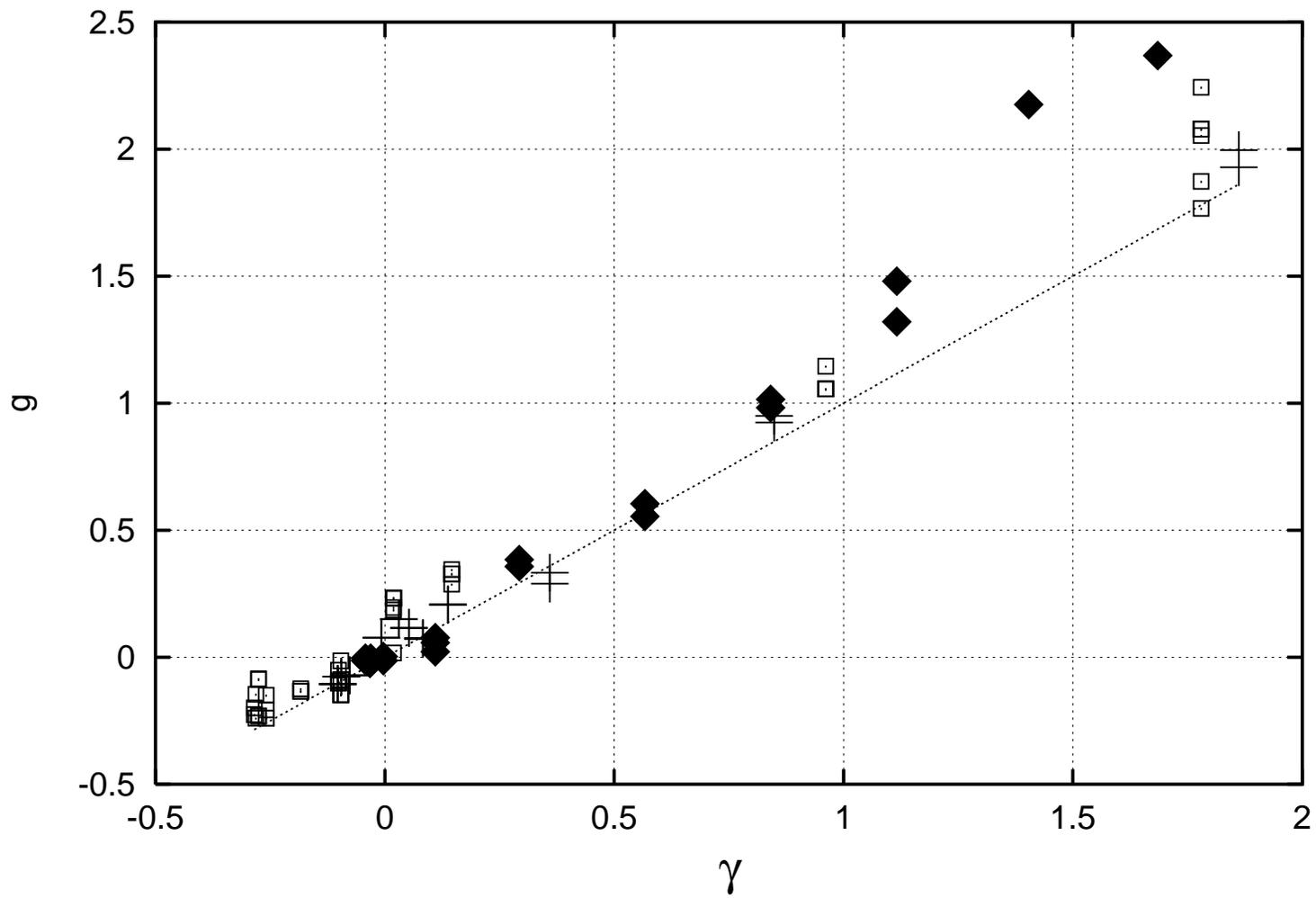
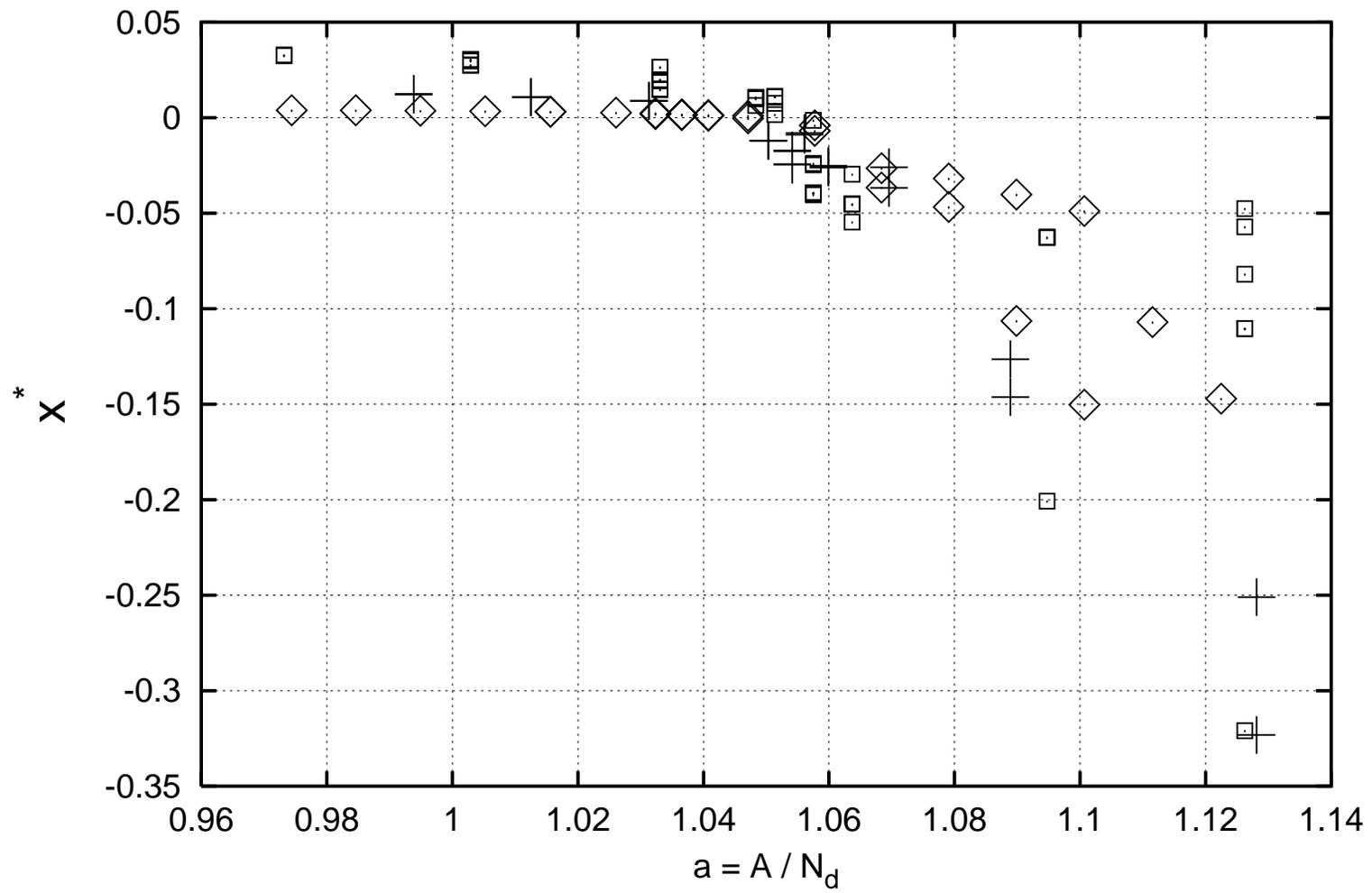
\end